\documentclass{article}
\usepackage{latexsym}
\usepackage{graphics}
\begin{document}
\hyphenation{Chri-sto-dou-lou}
\newcommand{\tl}{\frac{\theta}{\lambda}}
\newcommand{\zn}{\frac{\zeta}{\nu}}
\newtheorem{theorem}{Theorem}
\bibliographystyle{plain}
\title{Stability and Instability of the Reissner-Nordstr\"om Cauchy Horizon
and the Problem of Uniqueness in General Relativity}
\author{Mihalis Dafermos}
\date{\today}
\maketitle
\begin{abstract}
This talk will describe some recent results~\cite{md:si}
regarding the problem of uniqueness in the large (also known
as \emph{strong cosmic censorship}) for the
initial value problem in general relativity. The interest in
the issue of uniqueness in this context
stems from its relation to the validity of the
principle of determinism in classical physics.
As will be clear from below, this problem does not really have an analogue
in other equations of evolution typically studied. Moreover,
in order to isolate the essential analytic features of the
problem from the complicated setting of gravitational collapse 
in which it arises, some familiarity with the conformal properties of 
certain celebrated special solutions 
of the theory of relativity will have to be developed.
This talk is an attempt to present precisely these features to 
an audience of non-specialists, in a way which hopefully will 
fully motivate a certain characteristic initial value 
problem for the spherically-symmetric 
Einstein-Maxwell-Scalar Field system. 
The considerations outlined here leading 
to this particular initial value problem are well known in the physics 
relativity community, where 
the problem of uniqueness
has been studied heuristically~\cite{bdim:scsbhi,ispo:isbh} and 
numerically~\cite{brsm:bhs, bo:haifa}. 
In~\cite{md:si}, the global behavior of generic
solutions to this IVP, and in particular, the
issue of uniqueness, is completely understood. 
Only a sketch of the ideas of the proof
is provided here, but the reader may refer to~\cite{md:si} for details.
\end{abstract}

\section{General Relativity and its Initial Value Problem}
The general theory of relativity is thought to provide the correct 
classical description for the interaction of gravity with matter. 
This description is embodied in a system of partial
differential equations on a four dimensional manifold
$M$, the so-called \emph{Einstein equations}, 
which relate the Ricci curvature $R_{\mu\nu}$
of an unknown metric $g_{\mu\nu}$ to the energy-momentum 
tensor $T_{\mu\nu}$
of matter: 
\begin{equation}
\label{Einstein}
R_{\mu\nu}-\frac12Rg_{\mu\nu}=2T_{\mu\nu}.
\end{equation}
To complete the classical picture of a physics based on a collection of
fields satisfying a closed system of equations, one must also consider 
the laws which govern the evolution of the matter fields generating
the energy-momentum tensor on the right hand side of $(\ref{Einstein})$.
(One important special case is when there is no matter, the so-called
vacuum. Then $(\ref{Einstein})$ with vanishing right hand side
is a closed system of quasilinear hyperbolic equations.) In general,
one arrives at quite complicated systems
of equations. However, from the
perspective of classical physics, all phenomena are in 
principle described by the solutions of such
 a system. Moreover, for these systems, the initial 
value problem is natural, just as
in classical dynamics.

Thus, from one point of view, the general theory of relativity
is a classical physical theory that can be studied mathematically
in parallel with other
field theories of nineteenth-century classical physics.
Indeed, the equations of general relativity exhibit similar
local behavior with other equations of evolution as regards,
for instance, the issues of local existence and uniqueness
of solutions to the initial value problem.
When one turns, however, to
the initial value problem in the large, 
the Einstein equations present features that have 
no analogue in other typical equations of mathematical physics. 
The subject of this talk will be what appears, at least
at first sight, as the most pathological of these features,
namely the 
possibility of loss of uniqueness of the solution of 
the initial value problem \emph{without loss of regularity.} This
possibility is at the center of what is known as the
\emph{strong cosmic censorship conjecture} formulated by
Penrose~\cite{rp:gcrgr}.

The reason why the theory of the initial value problem in the 
large for the Einstein equations is richer than for other 
non-linear wave equations is that the global geometry of the 
characteristics is not constrained \emph{a priori} by any 
other structure. This geometry, which corresponds precisely to the 
conformal geometry for the vacuum 
equations, is \emph{a priori} unknown. It turns out that
many features of the initial value problem for
hyperbolic equations
that one takes for granted actually depend
on certain global properties of the geometry 
of the characteristics; the question of uniqueness
indicated above is one of these.

The best way to gain some intuition for what kind of 
conformal geometric structure develops in the course of 
evolution in general relativity--and what are the implications of
this structure--is to carefully examine the special 
solutions of the theory. In fact, almost all conjectures 
and intuition regarding the theory in the end derives from 
simple properties of such solutions. Moreover, since our focus
of interest is
global \emph{geometric} structure, there is
no substitute in building intuition than a good pictorial representation.
This talk will rely very much on such ``pictures''. It should be noted,
however, that in the spherically symmetric context in which we shall
be working, these ``pictures'', besides
conveying intuition, also carry complete
and precise information and can be treated on the
same level as symbols or formulas.

The assumption of spherical symmetry and associated
pictorial representations
will be carefully discussed
in the next section. We will then proceed to examine a series
of special solutions which will lead to a particular initial value
problem. Finally, theorems describing the solutions of the initial value
problem will be formulated and their proofs will be discussed. 

In regard to uniqueness, it turns out that
there is always a spacetime which can be uniquely associated to initial
data\footnote{For the vacuum, initial data is a Riemannian $3$-manifold
$(M,\tilde{g})$, along with a symmetric $2$-tensor $K$ satisfying the
constraint equations that would arise if $K$ were to be the
second fundamental form of $M$ realized as a hypersurface in a Ricci
flat $4$-manifold.}. 
This is the so-called \emph{maximal domain of development}~\cite{chge:givp}.
It is the ``biggest'' spacetime which admits the given
initial hypersurface as initial data
and is at the same time \emph{globally hyperbolic}, i.e.~all
inextendible causal curves intersect the initial hypersurface precisely
once. This latter property ensures that the domain of dependence
property holds. The question of uniqueness in general relativity
is thus the issue of the \emph{extendiblity} of this maximal
domain of development. If it is extendible, then the solution is
not unique. Since, as noted earlier, there is really no substitute
for a pictorial representation, we defer further discussion of
this till later on.

\section{Spherical Symmetry}
The current state of affairs in the theory of quasilinear 
hyperbolic partial differential equations in several space variables
is such that global, large data problems appear beyond reach. 
For there to be any hope of making headway, 
it seems that some sort of reduction must be made to a problem
where the number
of the independent variables is no more than two. For hyperbolic
equations of evolution, such reductions in general are accomplished
by considering symmetric solutions, or equivalently, symmetric
initial data. In general 
relativity, symmetry assumptions are formulated in terms of a 
group which acts by isometry on the spacetime and preserves all matter.

The only $2$-dimensional symmetry group that is compatible with 
the notion of an isolated gravitating system, i.e.~that can 
act on asymptotically flat spacetimes, is $SO(3)$. Solutions 
invariant under such an action are called \emph{spherically symmetric}. 
As we shall see below, most of the expected phenomena of
gravitational collapse of 
isolated gravitating systems, and the fundamental questions
that these phenomena pose,
can be suggested by the 
spherically symmetric solutions of various Einstein-matter 
systems. Moreover, the conformal structure of these solutions, 
which is the essential ingredient for the 
phenomena we wish to discuss, can be completely represented 
on the blackboard. (Or on paper!) The reason for this is 
simple: The space of group orbits \[Q=M/{SO(3)}\] can be given 
the structure of a $2$-dimensional Lorentzian manifold. 
Restricting to $Q$ which are
\emph{maximal domains of development} of initial 
data, it follows that these can be globally
conformally represented as bounded 
domains in $1+1$-dimensional Minkowski space. The images of
such representations are called \emph{Penrose diagrams};
from these, the conformal geometry can be immediately read 
off as the characteristics are just the lines at $\pi/4$ or $-\pi/4$ 
radians from the horizontal:
\[
\includegraphics{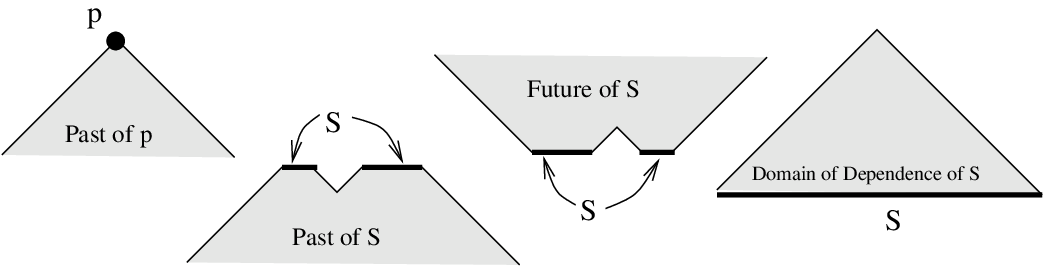}
\]

\section{Minkowski space}
To gain some familiarity with these diagrams, it is perhaps best to 
begin with $4$-dimensional Minkowski space from this point of
view, i.e.~with the Penrose diagram of the maximal domain of development
of Minkowski initial data. Here
the diagram is as follows:
\[
\includegraphics{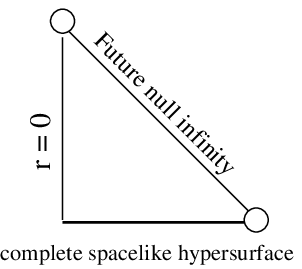}
\]
The $r$ referred to above is a function on $Q$ defined to be a multiple
of the square root of the area of the group orbit corresponding
to the points of $Q$. The line labelled $r=0$ is thus the axis of symmetry.
The line labelled ``future null infinity" is not part of the spacetime
but should be thought of as a ``boundary'' at infinity. The same 
applies to its
two endpoints, ``spacelike infinity'' and ``future timelike infinity''.
The latter corresponds to the ``endpoint'' of all 
inextendible future timelike geodesics.

The above Minkowski space is of course future causally 
geodesically complete, i.e.~all causal curves can be 
extended to infinite affine parameter. Thus we have the analogue
of global existence and uniqueness. That these properties 
of Minkowski space are stable to small
perturbations (\emph{without} any symmetry assumptions)
is a deep theorem of Christodoulou and Klainerman~\cite{ck:sms}.

\section{Schwarzschild}
Having understood the conformal diagram of Minkowski space, we turn to
a more interesting solution: the Schwarzschild solution. This is 
actually a one-parameter family of solutions 
(the parameter is called mass and denoted by $m$) 
which contains Minkowski space (the case where $m=0$). 
As it is a spherically symmetric vacuum solution, 
its non-triviality in the case 
$m\ne0$ must be generated by topology. Any
Cauchy hypersurface has two asymptotically 
flat ends and topology $S^2\times R$. ``Downstairs'', 
this corresponds to a line with two $r=\infty$ endpoints, 
not intersecting an axis of symmetry. For 
convenience, we will choose a time-symmetric initial 
hypersurface. The maximal development of this ``Schwarzschild'' 
initial data then looks like:
\[
\includegraphics{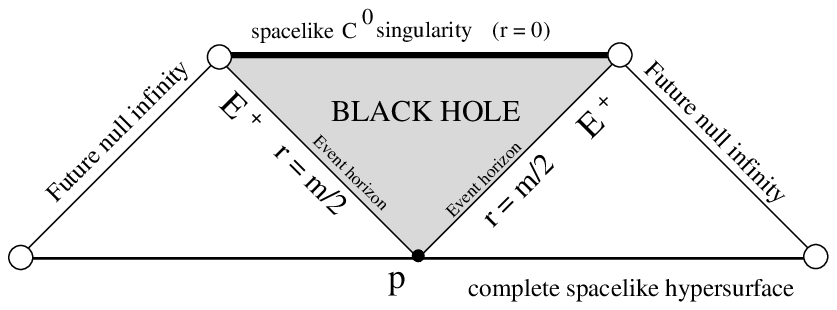}
\]
The point $p$ depicted where $r=\frac{m}2$ is a minimal surface
``upstairs''
in the initial hypersurface.
As the initial hypersurface was chosen to be time symmetric, 
this minimal surface is also what is called marginally trapped.
The ``ingoing''
and ``outgoing'' null cones emanating from this surface, 
which correspond to the two null rays through $p$ ``downstairs'',
 can thus not reach ``future null infinity''.
(This is related to the so-called ``singularity theorems'' of
Penrose.) 
Moreover, all timelike geodesics emanating from the point $p$ 
reach in finite time the curve $r=0$. This curve is not an axis 
of symmetry but a $C^0$ singularity! That is to say, 
there is no extension 
of the spacetime above through $r=0$ with a continuous Lorentzian metric.

This singular behavior of the Schwarzschild solution may at 
first appear to be an undesirable feature. Indeed, historically, 
it was first considered exactly as such. But it turns out in 
fact that the behavior outlined above would provide
the ``ideal'' scenario for 
the end-state of gravitational collapse. This has to do with 
two specific features of this solution:
\begin{enumerate}
\item
Labelling the null rays emanating from the minimal surface 
as $E^+$, it turns out that there is a class of timelike
observers, namely those who do not cross $E^+$, who can 
observe for infinite time and whose causal past is completely regular. 
The singularity is hidden inside a black hole, and $E^+$ is called
the event horizon. To be more precise about the ``completeness'' property
of the region outside the black hole,
fix some outgoing null geodesic which intersects future null 
infinity and parallel translate a conjugate null vector (i.e. a null vector in the other direction). The
affine length of the null curves joining this null geodesic 
and $E^+$, where the affine parameter is determined by the aforementioned
vector, goes to $\infty$. In particular, future null 
infinity can be thought to have infinite affine length. One
says that the solution pocesses a ``past complete future null infinity'' 
or a complete domain of outer communications.

\item
The above spacetime is
future inextendible as a $C^0$ metric. 
Thus, according to the discussion in
the Introduction, this means that
the Schwarzschild solution is unique even in the class
of very low regularity solutions. The significance of this fact
will become clear later.
\end{enumerate}

\section{Christodoulou's solutions}
At this point it should be noted that while the 
Schwarzschild solution indeed provides intuition 
about black holes, it cannot give insight as to 
whether these can occur in evolution of data where 
no trapped surfaces are present initially, i.e.~whether 
the kind of behavior outlined above is related in any 
way to the endstate of gravitational collapse. 
That Properties 1 and 2 above are indeed general properties of solutions
was proven by Christodoulou~\cite{chr:ins} for the spherically-symmetric 
Einstein scalar field equations:
\[
R_{\mu\nu}-\frac{1}{2}Rg_{\mu\nu}=2T_{\mu\nu},
\]
\[
g^{\mu\nu}(\partial_\mu\phi)_{;\nu}=0,
\]
\[
T_{\mu\nu}=\partial_\mu\phi\partial_\nu\phi-\frac12g_{\mu\nu}g^{\rho\sigma}
\partial_\rho\phi\partial_\sigma\phi.
\]
For generic solutions of the initial value problem,
the Penrose diagram obtained by Christodoulou is as follows:
\[
\includegraphics{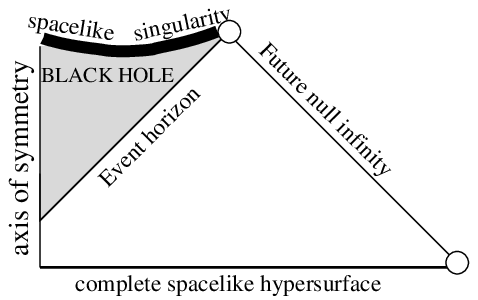}
\]
In~\cite{chr:en}, however, Christodoulou explicitly constructs
solutions 
with conformal diagram:
\[
\includegraphics{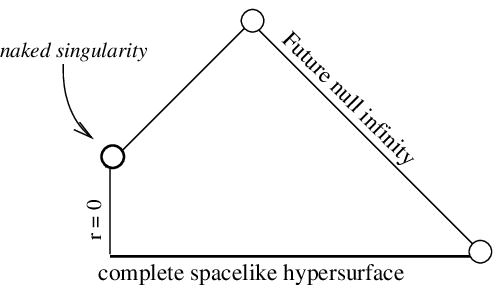}
\]
These are so-called ``naked singularities''.

One might ask why should one
consider the coupling with the scalar field.
The natural case to consider first, it would seem, is the
vacuum. A classical theorem of Birkhoff, however, states 
that the only spherically symmetric vacuum 
solutions are Schwarzschild. Thus, matter \emph{must} be included
to give the problem enough dynamical degrees of 
freedom in spherical symmetry. The scalar field 
is in some sense the simplest, most natural 
choice.\footnote{It satisfies a linear equation
and does not form singularities in the
absense of coupling, it is hyperbolic so does not change the
hyperbolic character of the equations, and moreover, 
its characteristics coincide with those of the metric $g$, etc\ldots}

As far as Property 1 is concerned, Christodoulou's results
are the best evidence yet that this is indeed a property
of ``realistic'' graviational collapse.\footnote{The statement
that for generic initial data, the domain of outer communications
is complete is known as ``weak cosmic censorship''.}
It turns out however that there is another ``competing'' set of
evidence that indicates that the behavior of Christodoulou's
solutions related to Property 2 does not represent ``realistic
collapse''. (Remember that Property 2 is the central question for
us, as this is what determines the notion of uniqueness.) This 
evidence is provided again by the intuition given by special solutions.

\section{The Kerr and Reissner-Nordstr\"om families}
One may consider the Schwarzschild family of solutions as 
embedded in a larger, 2-parameter family of solutions 
called the \emph{Kerr solutions}. Here the parameters are called
mass and angular momentum, and Schwarzschild corresponds 
to vanishing angular momentum. For all non-vanishing 
values of angular momentum, the internal structure of the 
black hole is completely different, and, as we shall see momentarily,
much more ``problematic'', as Property 2 will fail. Thus the
introduction of even an arbitrarily small amount
of angular momentum--a phenomenon that cannot be ``seen''
by spherically-symmetric models--seems to change everything
and cast doubt on the conclusions derived
from the spherically symmetric Einstein-Scalar Field
model.

To summarize our ``unhappy''
situation, it seems that the phenomenon which plays a fundamental role
in the issue we want to study is incompatible with the assumptions
we have to make in order to render it mathematically tractable. It would
seem that
understanding the black hole region of realistic gravitational
collapse using a spherically symmetric model is a lost cause.

Fortunately, there is a ``solution'' to this problem! The effect
of angular momentum on gravity turns out to be similar to the effect
of charge. (As John Wheeler puts it, charge is a poor man's angular
momentum.) Indeed, there is a very close similarity
between the conformal structure of
the Kerr family and a 2-parameter \emph{spherically symmetric}
family of solutions to the
Einstein-Maxwell equations: 
\[
R_{\mu\nu}-\frac{1}{2}g_{\mu\nu}R=2T_{\mu\nu}
\]
\[
F^{\mu\nu}_{;\nu}=0,
\]
\[
F_{[\mu\nu,\rho]}=0,
\]
\[
T{\mu\nu}=F_{\mu\lambda}F_{\nu\rho}g^{\lambda\rho}
-\frac1{4}g_{\mu\nu}F_{\lambda\rho}F_{\sigma\tau}
g^{\lambda\sigma}g^{\rho\tau},
\]
the so-called \emph{Reissner-Nordstr\"om}
solution. Here the parameters are mass $m$ and charge $e$. 
For $e=0$ one retrieves the Schwarzschild family, while for
$0<e<m$ one obtains:
\[
\includegraphics{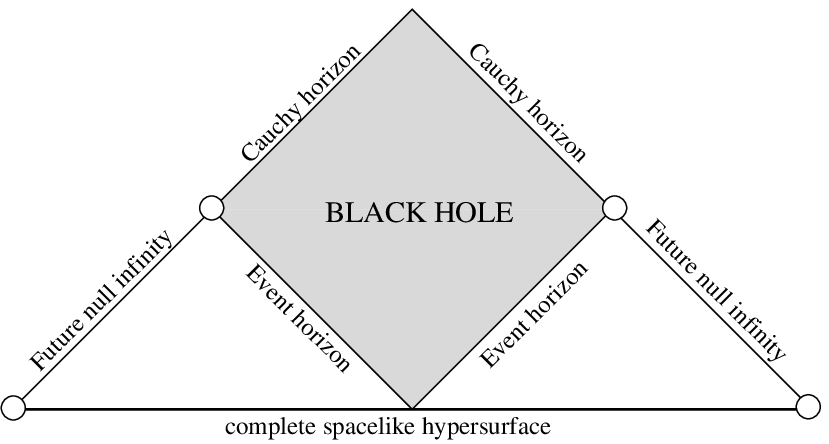}
\]
The $r=0$ singularity of the Schwarzschild solution has
disappeared! The above spacetime is completely regular up to the
edges. These new
edges that ``complete'' the triangle, 
however, are at a \emph{finite} distance from
the initial data, in the sense that all timelike geodesics
joining those edges with the initial hypersurface have finite length.
This solution thus has a regular future boundary and is
extendible (in $C^\infty$!) beyond it.

What fails at the boundary
of this maximal domain of development of initial data is thus not
the regularity of the solution, but rather,
global hyperbolicity. Any extension
of $Q$ will contain past inextendible causal geodesics not intersecting
the intial hypersurface:
\[
\includegraphics{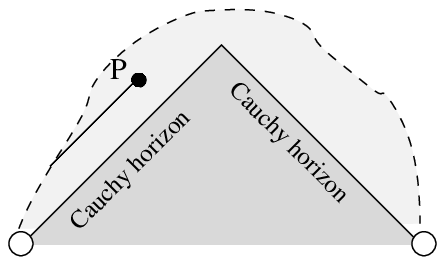}
\]
Points in such an extension but not in $Q$ itself
cannot be determined by initial data, in the
exact same way that the solution to a linear
wave
equation $\Box\Psi=0$ at the point $P$ depicted below, cannot be
uniquely determined by its values in the shaded set $S$:
\[
\includegraphics{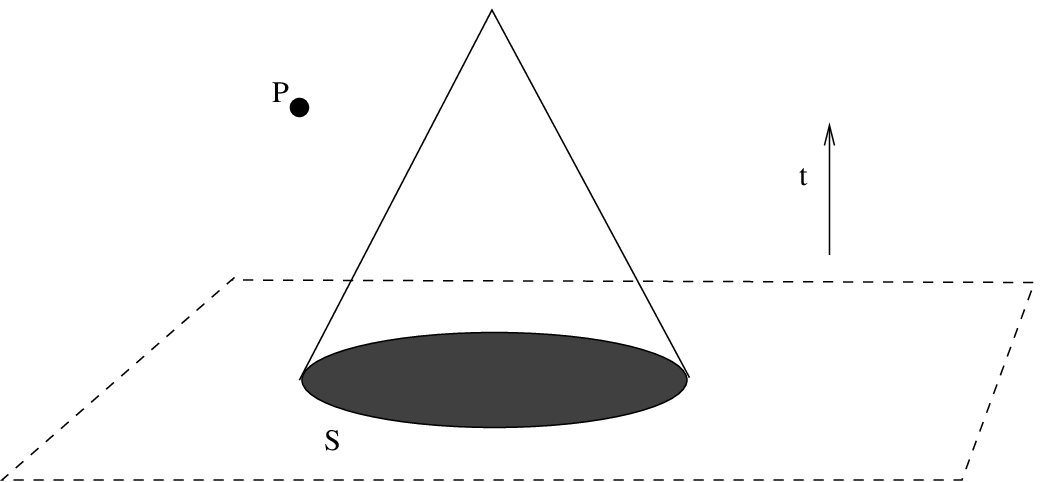}
\]
What has happened in the Reissner-Nordstr\"om solution is
that the situation depicted above developed, but from
an $S$ that was complete. 

\section{Strong Cosmic Censorship}

The physical interpretation of the above situation is that
the classical principle of determinism fails, but without any sort
of loss in regularity that would indicate that the domain of the
classical theory has been exited. It is in that sense that
this kind of behavior is widely considered by physicists
to be problematic\footnote{Of course, this failure
of determinism only applies to observers who enter into black holes. 
In particular, Property 1 ensures that determinism holds in the
domain of outer communications (hence the term ``weak'' cosmic 
censorship). On the other hand, it seems that fundamental principles
of physics should be valid everywhere, 
including the interiors of black holes.}. On the other hand,
numerical calculations (Penrose
and Simpson~\cite{sp:RN}) on the
behavior of linear equations on a Reissner-Nordstr\"om background
indicate that a naturally defined derivative 
blows up at the Cauchy horizon. This was termed the blue-shift
effect. Thus, Penrose argued,
the pathological behavior of the Reissner-Nordstr\"om solution
might be unstable to perturbation. 
This led him to conjecture, more generally, 
\vskip1pc
\noindent\emph{Strong Cosmic Censorship} For generic initial data,
in an appropriate class,
the maximal domain of development is inextendible.
\vskip1pc
In view of our discussion in the introduction, this can be thought
of as the conjecture that, for generic initial data, the solution
is unique wherever it can be defined. Of course, the context in which
this should be applied to (i.e.~what equations, what class of initial
data should be considered, etc\ldots)
and the notion of extendibility are left open. We will comment more
on that later.

\section{The Einstein-Maxwell-Scalar Field system}
It might seem at first
that the proper setting for discussing the problem of whether
Cauchy horizons arise in evolution, for generic data in spherical
symmetry, is the Einstein-Maxwell equations. 
Unfortunately, these suffer in fact from the same drawback
as the Einstein vacuum equations, namely they do not possess the
required dynamical degrees of freedom. One necessarily has to include
more matter, and again the simplest choice, as in the work
of Christodoulou, is a scalar field. Thus one is easily led to
the coupled Einstein-Maxwell-Scalar Field system:

\[
R_{\mu\nu}-\frac{1}{2}g_{\mu\nu}R=2T_{\mu\nu}=
2(T_{\mu\nu}^{em}+T_{\mu\nu}^{sf})
\]
\[
F^{\mu\nu}_{;\nu}=0,
\]
\[
F_{[\mu\nu,\rho]}=0,
\]
\[
g^{\mu\nu}(\partial_\mu\phi)_{;\nu}=0,
\]
\[
T^{em}_{\mu\nu}=F_{\mu\lambda}F_{\nu\rho}g^{\lambda\rho}
-\frac1{4}g_{\mu\nu}F_{\lambda\rho}F_{\sigma\tau}
g^{\lambda\sigma}g^{\rho\tau},
\]
\[
T^{sf}_{\mu\nu}=\partial_\mu\phi
\partial_\nu\phi-\frac12g_{\mu\nu}g^{\rho\sigma}
\partial_\rho\phi\partial_\sigma\phi.
\]
It turns out that in spherical symmetry the Maxwell part
of the equation decouples. Since the scalar field $\phi$ carries no charge,
a non-trivial Maxwell field can only be present if an initial complete
spacelike hypersurface
has non-trivial topology. In particular, this model is not suitable
for considering the formation of black holes, as in the work of 
Christodoulou. Thus we will consider
the problem where there is already a black hole present initially.
 To take the simplest possible
formulation that captures the essense of the problem at hand, 
one can prescribe initial data for the system on two null rays,
such that one corresponds to the event horizon of a Reissner-Nordstr\"om
solution, and the other carries ``arbitrary'' matching data:
\[
\includegraphics{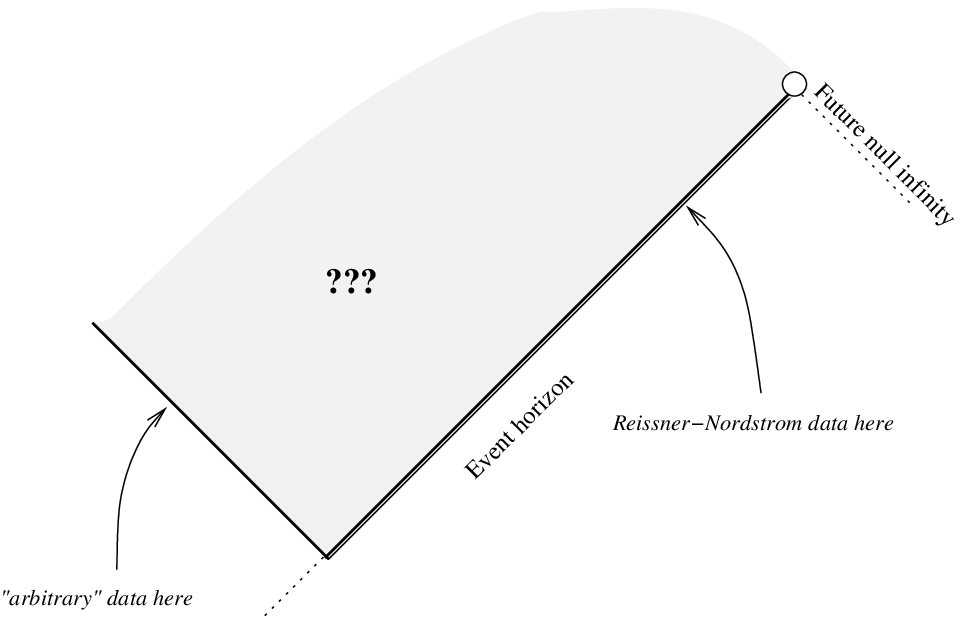}
\]
To make the most of the method of characteristics, we introduce
null coordinates, i.e.~coordinates $(u,v)$ such that the metric on $Q$
takes the form $-\Omega^2dudv$. Here we select the $v$-axis to be
the event horizon, and the $u$ axis to be the conjugate ray on which
we presribe our data. The unknowns are then just $r$, $\Omega$
and $\phi$, and the electromagnetic part contributes a constant $e\ne0$
which is computed from the initial data.
To write the equations as a first order system, we define
$\partial_ur=\nu$, $\partial_vr=\lambda$, $r\partial_u\phi=\zeta$,
$r\partial_vr=\zeta$,
and also $\varpi$, what one can call the ``renormalized'' Hawking 
mass\footnote{The Hawking mass $m$ is defined to be
$m=\frac{r}2(1-|\nabla r|^2)$. The renormalized version has
the property that it is constant in the Reissner-Nordsr\"om solution
and coincides with $m$ at Future Null Infinity.},
defined by
\[
1-\frac{2\varpi}r+\frac{e^2}{r^2}=|\nabla r|^2=-4\Omega^{-2}\lambda\nu.
\]
We then have:
\begin{equation}
\label{ruqu}
\partial_u{r}=\nu,
\end{equation}
\begin{equation}
\label{rvqu}
\partial_v{r}=\lambda,
\end{equation}
\begin{equation}
\label{nqu}
\partial_v\nu=\nu\left(-\frac{2\lambda}{1-\mu}\frac{1}{r^2}
\left(\frac{e^2}{r}-\varpi\right)\right), 
\end{equation}
\begin{equation}
\label{puqu}
\partial_u\varpi=\frac{1}{2}(1-\mu)\left(\zn\right)^2\nu,
\end{equation}
\begin{equation}
\label{pvqu}
\partial_v\varpi=\frac{1}{2}(1-\mu)\left(\tl\right)^2\lambda,
\end{equation}
\begin{equation}
\label{sign1}
\partial_u\theta=-\frac{\zeta\lambda}r,
\end{equation}
\begin{equation}
\label{sign2}
\partial_v\zeta=-\frac{\theta\nu}r.
\end{equation}

\section{Statement of the theorems}
We can now state the theorems of~\cite{md:si}. On the one hand we have:
\begin{theorem} 
After restricting the range of the $u$ coordinate,
the Penrose diagram of the solution of the I.V.P.~described
above is as follows:
\[
\includegraphics{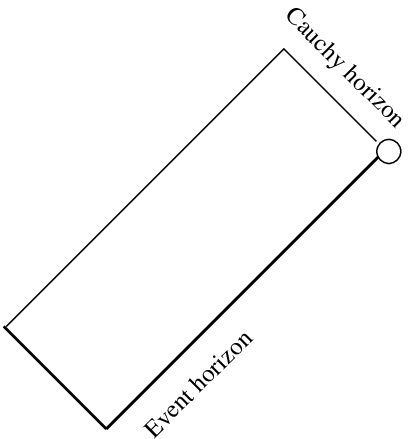}
\]
Moreover, $r$ extends to a function on the Cauchy horizon with
the property
$r\to \varpi_{init}-\sqrt{\varpi_{init}^2-e^2}$ as $u\to0$ (where 
$\varpi_{init}$ is the constant value of $\varpi$
on the initial \emph{event} horizon), 
and the metric
can be continuously extended globally across the Cauchy horizon.
\end{theorem}
On the other hand, we have:
\begin{theorem} 
For ``generic'' initial data in the class of allowed
initial data, $\varpi$ blows up identically on the Cauchy horizon.
In particular, the solution is inextendible across the Cauchy horizon
as a $C^1$ metric.
\end{theorem}

Thus, strong cosmic censorship is true, according to Theorem 2,
if formulated with respect to extendibility in $C^1$ or higher,
but false, according to Theorem 1, if formulated with respect to 
extendibility in $C^0$ (See~\cite{chr:givp} for reasons why one might
want to require $C^0$.) In any case,
the formation of a null ``weak'' singularity indicates
a qualitatively different picture of the internal structure
of the black hole from any of the previous models described
above, and from the original expectations of Penrose.

The scenario of Theorems 1 and 2 was
first suggested by Israel and Poisson~\cite{ispo:isbh} who put forth some
heuristic arguments. Subsequently, a large class of numerics
was done for precisely the equations considered here (see~\cite{bo:haifa}
for a survey). Because of the blow-up in the mass, 
this phenomenon was termed ``mass inflation''.

It should be noted that the imposition of Reissner-Nordstr\"om data
on the event horizon is somewhat
unnatural if the data is viewed 
as having arisen from generic data
for a characteristic value problem where the $v$-axis is
in the domain of outer communications\footnote{It should
be emphasized that our rationale in
choosing the I.V.P.~here was to separate completely the issue of the
dynamics
of the interior of the black hole from the 
precise understanding of the set
of data that arises in collapse, which is a problem of
very different analytical flavor.}. 
Similar results to the above theorems, however, can in fact be proven for 
a wide class of data which includes the kind conjectured to arise from
the aforementioned problem.
These results will appear in~\cite{md:si2}. The relavence of such
an extension will only become clear, however, if
the problem of determining the correct ``generic''
decay on the event horizon is mathematically resolved.

\section{Some ideas from the proofs}
Our initial data are trapped, i.e.~$\nu$ and $\lambda$ are both
nonpositive. These signs are then preserved in 
evolution. Note that from the equation
\begin{equation}
\label{lmqu}
\partial_u{\left(\frac{\lambda}{1-\mu}\right)}=
\left(\frac{\lambda}{1-\mu}\right)\frac{1}r\left(\zn\right)^2\nu
\end{equation}
it follows that
$\frac{\lambda}{1-\mu}$ is non-increasing in $u$\footnote{That this quantity
is non-increasing is in fact a general feature of spherical symmetry,
i.e.~it depends only on the dominant energy condition, not on
the particular choice of matter.}. 
What gives the analysis of our equations \emph{in the
 black hole region} its characteristic flavor 
is the fact
that $\int{\frac{\lambda}{1-\mu}}dv$ is potentially infinite when 
integrated for constant $u$ along the whole range of $v$ (it
is indeed infinite in initial data, i.e.~$u=0$), and that this infinity can 
appear in the equations (for instance in $(\ref{nqu})$)
with either a positive or negative
sign, depending on the sign of $\frac{e^2}r-\varpi$.
Note that, by contrast, in the domain of outer communications, 
this infinity is killed by the $\frac1{r^2}$ term since 
$r\to\infty$ on outgoing rays.
In the domain of development of our initial data, $r$ is
bounded above by its initial constant value on the event
horizon $\varpi_{init}+\sqrt{\varpi_{init}^2-e^2}$, in view
of the signs of $\lambda$ and $\nu$.

In the Reissner-Nordstr\"om solution, the sign 
of $\frac{e^2}r-\varpi$ goes
from negative near the event horizon to positive near the Cauchy
horizon, while $\int{\frac{\lambda}{1-\mu}dv}$ remains constant in $u$
and thus infinite, accounting for both what is called
the infinite red shift near the event horizon (this makes objects
crossing the event horizon slowly disappear to outside observers
as they are shifted to the red) and the infinite blue shift
near the Cauchy horizon (this accounts for the instability of
the Cauchy horizon to \emph{linear} perturbations).

For general solutions of the initial value problem, some
of these features of the sign of $\frac{e^2}r-\varpi$ turn out to be stable,
while others do not. In particular, Theorems 1 and 2 together
imply that the sign must become negative near the Cauchy horizon,
and not positive! To attack this initial value problem, it is
clear that the behavior of this sign
is the first thing that must be understood.
It turns out that before the effects of the linear instability 
start to play a role, three geometrically
distinct regions develop in evolution, a red-shift, no-shift,
and stable blue-shift region, characterized by
\[
\frac{e^2}r-\varpi<-\epsilon{\rm\ ,\ }\frac{e^2}r-\varpi\sim0
{\rm\ ,\ }
\frac{e^2}r-\varpi>\epsilon,
\]
respectively:
\[
\includegraphics{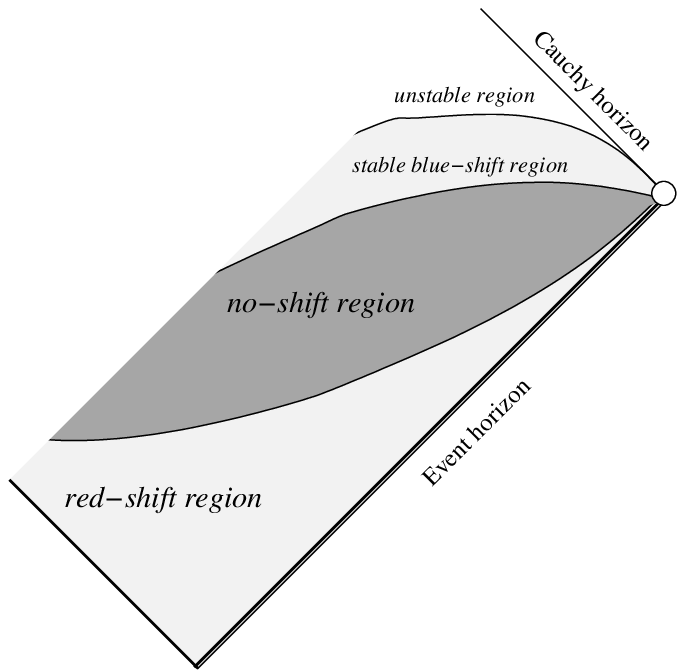}
\]
In the red-shift region, $\int{\frac{\lambda}{1-\mu}}dv$ is unbounded
as $u\to0$,
but it appears with a ``favorable'' sign (favorable as far as 
controlling $\varpi$ is concerned\footnote{This sign tends to make
$\nu$ bigger, and there is an extra $\nu$ on the denominator of
$(\ref{puqu})$.}), in the no-shift region,
$\int{\frac{\lambda}{1-\mu}}$ is uniformly bounded, and in
the stable blue-shift region, $\int{\frac{\lambda}{1-\mu}}dv$
grows with $u$ but at a rate ``less'' than the growth of certain
natural derivative of $\phi$. These facts allow us to control
all quantities reasonably well up until the future boundary
of the stable blue-shift region, though completely different
arguments must be applied to each subsequent region as it
develops in evolution from the previous one.

Of course, all this work seems only to have pushed forward the
problem from the original initial segments to the future boundary
of
the ``stable blue-shift region''.
But in fact
our new ``initial'' conditions on the future boundary of the
stable blue-shift region are much more favorable. The stable
blue-shift region that has preceded it ensures that
$\nu$ has a sufficiently fast decay rate in $u$. (Remember that
blue-shift regions tend to make $|\nu|$ smaller, so they are favorable
for controlling $r$, but unfavorable for controlling $\varpi$.)
Once this rate can be shown to be preserved, it follows
by integrating $u$ that one can bound $r$ \emph{a priori}
away from $0$ in its future, and thus prove the existence
of the solution up to the Cauchy horizon. It is clear from what
we have said above that if the unstable region remains a blue-shift
region (see left diagram below), there is no problem. (This is of course
what happens for the Reissner-Nordsr\"om solution itself.)
\[
\includegraphics{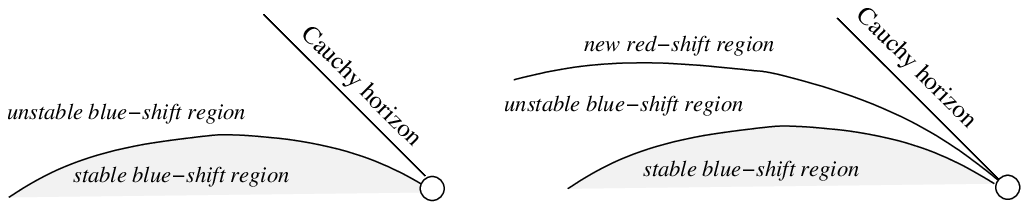}
\]
The danger is if a new red-shift region develops (see right diagram
above). It turns out, however, that a new \emph{a priori} estimate
\[
\int{\partial_v\log(\nu)dv}=-\int{\frac{2\lambda}{1-\mu}\frac{1}{r^2}
\left(\frac{e^2}{r}-\varpi\right)dv}<C
\]
is available in this case, which is
independent of the size of $\varpi$.
This makes use of the fact that there is a $\varpi$ hidden
also in the denominator
in $1-\mu$. (The estimate depends, however, on our knowledge
of the new initial condition of $r$ and a particular bootstrap
assumption on its future behavior; in particular this estimate
does not hold in the original ``red-shift'' region.) This is
the last element in the proof of Theorem 1. 

We now discuss
the proof of Theorem 2. As has been noted, at the level of perturbation
theory~\cite{ch:cchRN}, one does see an instability, caused by the 
blue-shift region.
(Requiring some derivative of the scalar field to be positive initially,
$\zeta$ will decay in $u$, and thus $\theta$ will decay in $v$, at
a slower rate than the decay of $\lambda$ in $v$, so that the natural
derivative $\tl\to\infty$.)
On the other hand, the non-linearity tends to diminish
this effect, since if the mass indeed increases, in view of Theorem 1,
we have a reappearance of a red-shift
region (see diagram on the right above). As remarked earlier,
this tends to make $|\nu|$ (and also $|\lambda|$) bigger,
and thus $\left|\tl\right|$ smaller. 
The proof must thus encorporate something beyond ``linear
theory arguments''. 

This extra ingredient is supplied by
a very powerful monotonicity peculiar to black hole interiors, or
more
specifically, to ``trapped regions''. Integration of $(\ref{sign1})$ and
$(\ref{sign2})$ then implies that if $\theta$ and $\zeta$ are
initially of the same signs\footnote{In view of the fact
that the data vanish on the event horizon, this can be considered
a generic condition after restricting the domain of the $u$ coordinate.
This condition and the non-vanishing of a particular derivative of $\phi$
at the origin together define the ``generic'' class of
initial data to which Theorem 2 applies.}, 
let's say non-negative,
then they remain non-negative, and in fact
$\partial_v\zeta\ge0$ and $\partial_u\theta\ge0$. 
Now integration of $(\ref{pvqu})$ and using
that $\frac{1-\mu}{\lambda}$ is also non-increasing in $u$,
yields that for $v_2>v_1$, $u_2>u_1$,
\begin{equation}
\label{monot}
\varpi(u_2,v_2)-\varpi(u_2,v_1)\ge\varpi(u_1,v_2)-\varpi(u_1,v_1).
\end{equation}
It should be mentioned that in view of the sign of $\lambda$,
both sides of the above inequality are positive.

The broad outline of the proof of Theorem 2 is as follows:
First assume that the spacetime looks very much like Reissner-Nordstr\"om.
Then the linear theory more or less applies, and applying the bounds for
$\theta$ in the equation $(\ref{pvqu})$, one obtains $\varpi\to\infty$,
which is a contradiction. Thus one is reduced to proving that any
spacetime ``quantitatively different'' from Reissner-Nordstr\"om
must have $\varpi\to\infty$.

It is not possible here to explain precisely
what ``quantitatively different'' has to mean. To give
a taste of the kind of arguments involved in this ``non-linear
part'' of the proof, we will be content to show
that assuming only that $\varpi$ is bounded below 
by a positive number plus its Reissner-Nordstr\"om value
(this is indeed a quantitative difference)
\begin{equation}
\label{assume}
\varpi_{Cauchy\ horizon}>\varpi_{init}+2\epsilon,
\end{equation}
it follows from
$(\ref{monot})$ that $\varpi$ must in fact
blow up identically on the Cauchy horizon. 
If $\gamma$ is the future boundary of the stable blue-shift
region, the
fact that $\varpi_\gamma(u)\to\varpi_{init}$ implies that
given any $u_0$,
a sequence of points $(u_i,v_i)$ can be constructed
so that the mass differences $\varpi(v_{i+1},u_i)-\varpi(v_i,u_i)$
are all greater than $\epsilon$:
\[
\includegraphics{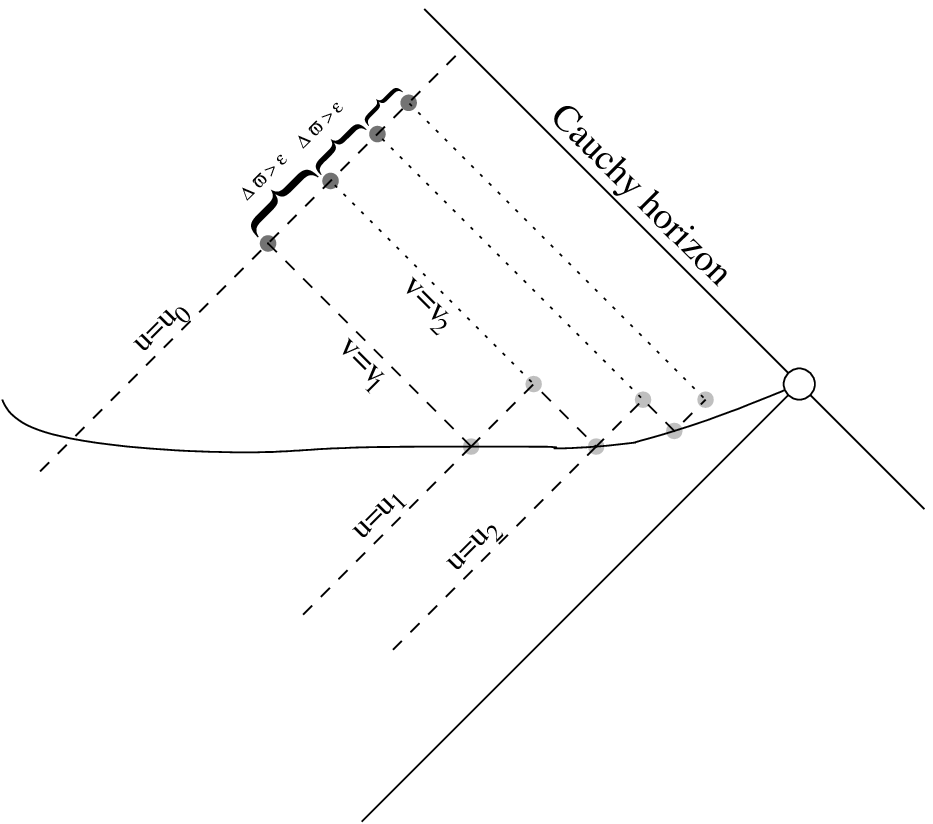}
\]
But in view of $(\ref{monot})$, this mass difference
can be added on $u=u_0$ to yield infinite mass at
the point where $u=u_0$ meets the Cauchy horizon.

Suffice it to say that the non-linear analysis of black hole
regions is quite different than the analysis we are used to.
Which parts of this spherically-symmetric picture
generalize and which do not remains to be seen.

\end{document}